\documentclass[prl,aps,twocolumn,superscriptaddress]{revtex4-2}

\usepackage{color}
\usepackage{bm}
\usepackage{units}
\usepackage{graphicx}
\usepackage{bbold}
\usepackage{tikz}
\usepackage{graphicx}
\usepackage{bm}
\usetikzlibrary{calc}
\usetikzlibrary{arrows}
\usepackage{hyperref}
\usepackage{comment}

\usepackage{physics}

\usepackage[justification=raggedright, singlelinecheck=false]{caption}
\usepackage{mathtools}
 
\usepackage[normalem]{ulem}

\usepackage{hyperref}
\makeatletter
\newcommand*{\rom}[1]{\expandafter\@slowromancap\romannumeral #1@}
\makeatother

\definecolor{indigo}{rgb}{0.44, 0.0, 1.0}

\definecolor{ywcolor}{rgb}{0.0, 0.5, 0.0}

\begin{document}

\title{Magnetic Order in bilayer Ruddlesden-Popper Nickelates}

\author{Yiming Wang}
\thanks{These authors contributed equally to this study.}
\affiliation{Department of Physics \& Astronomy,  Extreme Quantum Materials Alliance, Smalley-Curl Institute,
Rice University, Houston, Texas 77005, USA}

\author{Guijing Duan}
\thanks{These authors contributed equally to this study.}
\affiliation{School of Physics and Key Laboratory of Quantum State Construction and Manipulation (Ministry of Education),
Renmin University of China, Beijing, 100872, China}

\author{Zhiguang Liao}
\affiliation{School of Physics and Key Laboratory of Quantum State Construction and Manipulation (Ministry of Education),
Renmin University of China, Beijing, 100872, China}

\author{Kuan-Sen Lin}
\affiliation{Department of Physics \& Astronomy,  Extreme Quantum Materials Alliance, Smalley-Curl Institute,
Rice University, Houston, Texas 77005, USA}

\author{Rong Yu}
\email{rong.yu@ruc.edu.cn}
\affiliation{School of Physics and Key Laboratory of Quantum State Construction and Manipulation (Ministry of Education),
Renmin University of China, Beijing, 100872, China}

\author{Qimiao Si}
\email{qmsi@rice.edu.cn}
\affiliation{Department of Physics \& Astronomy,  Extreme Quantum Materials Alliance, Smalley-Curl Institute,
Rice University, Houston, Texas 77005, USA}

\begin{abstract}
The recent discovery of high-temperature superconductivity in the bilayer nickelate La$_3$Ni$_2$O$_7$ has led to extensive interest in the correlation physics of its normal state. 
Given that the superconducitivity develops near a density wave order in the phase diagram, it is important to elucidate the nature of this order. 
Based on the accumulated experimental evidence for a bad metal state in proximity to an orbital-selective Mott phase, 
here we describe magnetic correlations of the system in a conceptually new way -- in terms 
of effective local moments experiencing a combination of RKKY and superexchange interactions.
This gives rise to a magnetic order with a wavevector 
that is close to  $\mathbf{Q}=(\pi/2,\pi/2)$ and, at the same time,
yields a clear understanding of the associated spin dynamics. 
Our results are consistent with the rapidly emerging  experiments about the magnetic correlations 
in the density wave order of the bilayer nickelate. 
Implications for unconventional superconductivity
in this and related multiorbital systems are discussed.
\end{abstract}
\maketitle

\textit{Introduction.}~
Bilayer nickelate La$_3$Ni$_2$O$_7$
shows high temperature superconductivity 
both in bulk crystals under pressure~\cite{Sun_Nature_2023} and in thin films at ambient 
pressure~\cite{Hwang_Nature_2024,zhouAmbientpressureSuperconductivityOnset2025,Hwang_film2501}.
The discovery has 
motivated extensive experimental investigations
on this and related 
Ruddlesden-Popper nickelates~\cite{Yuan_NP_2024,arXiv:2407.05681Cheng,Hao2025,Zhou2025, Liu2025,Zhou2026,Fan2026,Plokhikh_arxiv2503.05287}.
It also opens a new window into the understanding of the basic physics of high temperature 
and unconventional supercondcutors and, as such, led to enormous theoretical
efforts~\cite{Liao_PRB_2023,QuSu_PRL_2024, HeierSavrasov_PRB_2024,
ZhanHu_arXiv_2024, TianLu_PRB_2024, ChangLi_arXiv_2023, QinYang_PRB_2023, LuoYao_npjQM_2024, JiangZhang_CPL_2024,
HuangZhou_PRB_2023, YangZhang_PRB_2023, LuWu_PRL_2024, XueWang_CPL_2024, ChenLi_PRB_2024, KanekoKuroki_PRB_2024, KakoiKuroki_PRB_2024,
SakakibaraKuroki_PRL_2024, YangWang_PRB_2023,
JiangKu_PRL_2024, LiuChen_arXiv_2023, QuSu_arXiv_2023,
YangZhang_arXiv_2023, ZhangWeng_PRL_2024, LuYou_arXiv_2023, FanXiang_PRB_2024,
ZhengWu_arXiv_2023, SchlomerBohrdt_arXiv_2023, BotzelEremin_arXiv_2024_1,
BotzelEremin_arXiv_2024_2, OhZhang_PRB_2023, ZhangDagotto_PRB_2023, ZhangDagotto_NC_2024,Mei_arxiv2605.20148,Duan2026}.
Importantly,
optical conductivity and angle resolved photoemission spectroscopy (ARPES) measurements
at ambient pressure have provided 
fairly direct evidence for the importance of
orbital-selective electron correlations
\cite{LiuWen_NC_2024,YangZhou_NC_2024}. 
These spectroscopic experiments are captured
in terms of bad metallicity with orbital-selective Mott correlations \cite{Liao_PRB_2026,Liao_PRB_2023,duan_arxiv2502.09195}. 
Combined with first-principles studies, it has become clear that the low-energy physics is 
primarily governed by Ni 3$d$ electrons in the $e_g$ manifold --
the strongly hybridizing 
$3d_{x^2-y^2}$ and $3d_{z^2}$ orbitals~\cite{daoxinyao2023}. 
The $d_{x^2-y^2}$ orbitals predominantly dictate the in-plane intralayer hopping, whereas the $d_{z^2}$ orbitals facilitate strong interlayer coupling, creating a highly anisotropic electronic environment.

While epitaxial strain enables superconductivity in thin films at ambient pressure, the bulk compound under ambient conditions is known to exhibit an electronic density-wave order 
at low temperatures (below about $150$ K).
We start from the salient experimental observations about this density-wave order 
at the ambient pressure.
Local probes
have played an integral role.
Zero-field muon spin resonance ($\mu$SR) measurements reported bulk magnetic order below $T_N\simeq 154$ K, with oxygen deficiency broadening the internal field distribution~\cite{Chen_muSR_2023}.  Consistently, $^{139}$La 
nuclear magnetic resonance (NMR) measurements found a 
magnetic transition around the same temperature~\cite{Zhao_NMR_2025}.    
More recently, $^{139}$La nuclear quadrupolar resonance (NQR) measurements
indicated that charge modulation and magnetic broadening appear together below $T_{\rm DW}\sim153$ K~\cite{Luo_NQR_2025}.  These results 
implicate the spin and charge 
correlations of the system at ambient conditions.

As a more direct motivation, recent scattering measurements have begun to clarify the structure and dynamics of the ordered state. 
Resonant inelastic x-ray spectroscopy (RIXS) in a bilayer nickelate observed well-defined magnetic excitations~\cite{Chen_RIXS_2024}, 
while resonant soft x-ray scattering has revealed a 
related magnetic order~\cite{guptaAnisotropicSpinStripe2025}.   
Neutron diffraction 
subsequently resolved 
the magnetic structure,
with magnetic scattering appearing below approximately 150 K and an antiferromagnetic stacking of the two layers within a bilayer~\cite{Plokhikh_ND_2025, Zhong2026}.  Most directly, recent inelastic neutron scattering 
measurements on single-crystalline La$_3$Ni$_2$O$_7$ observed spin excitations centered near $
(0,0.5,2.5)$ 
[corresponding to an in-plane wavevector 
$(\pi/2, \pi/2)$ in the unfolded 1-Ni Brillouin zone]~\cite{Chen_INS_2026}.

It is believed that the density-wave order comes from the same
electronic degrees of freedom (the aforementioned 
Ni 3$d$-electrons in the $e_g$ manifold)
that are responsible for the correlated normal state and superconductivity.
Accordingly,
understanding the nature of the density wave 
order 
is 
a key to
elucidating the microscopic physics of this high temperature superconductor.
The theoretical approach to this issue 
is still limited, mostly from 
a weak coupling Fermi surface nesting instabilities
\cite{fanyang2025, Zhan_arxiv2606.20533, Braz_arxiv2606.29527}
or 
the strong coupling limit with purely localized 
moments 
experiencing 
superexchange interactions \cite{Ni_arxiv2407.19213, wang2026originspinstripesbilayer}.
How to approach this problem building on the understanding about 
the normal state remains an open issue.

\begin{figure}
\centering
     \includegraphics[width=\linewidth]{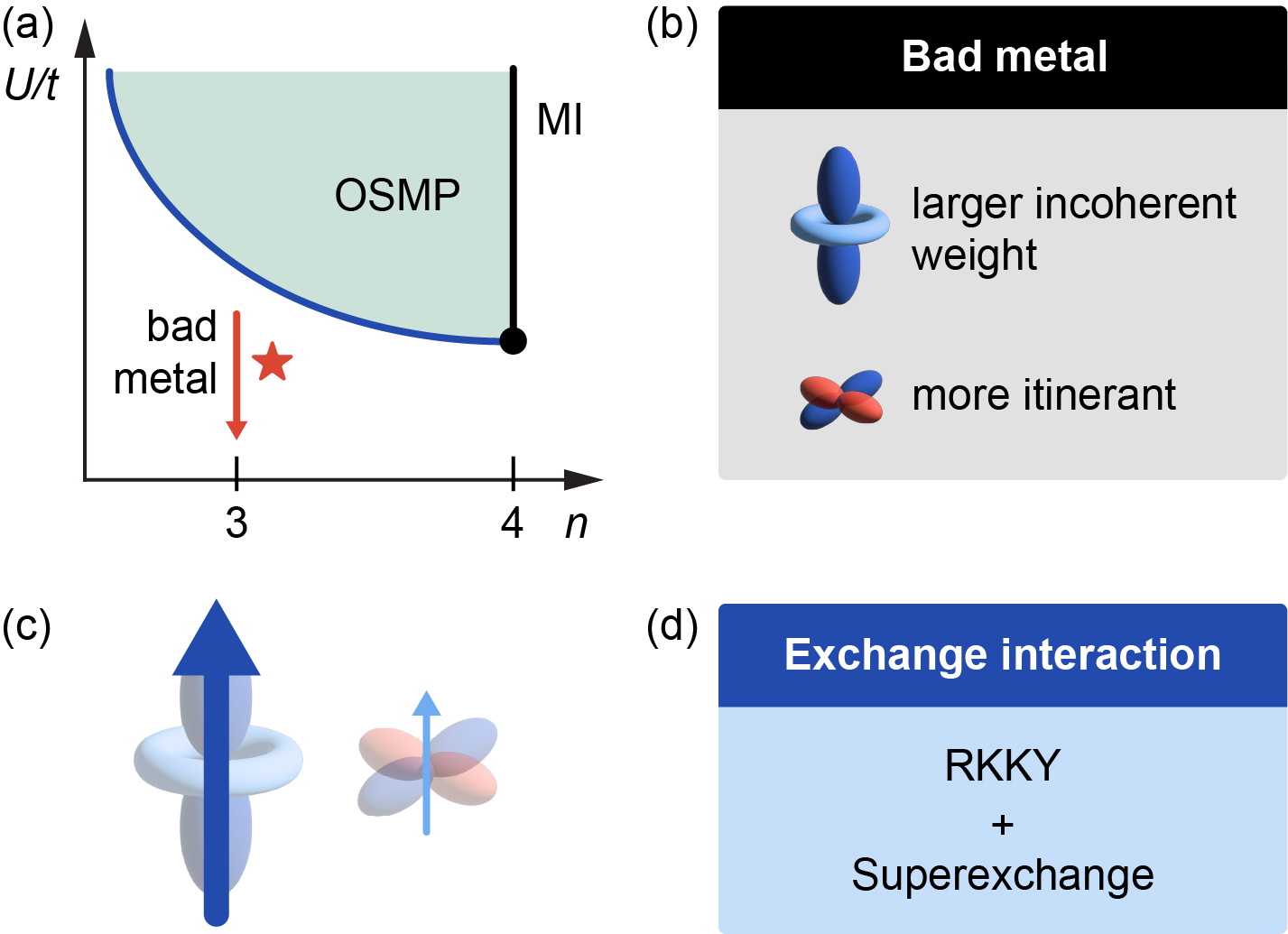}
    \caption{   
    (a) Sketched ground-state phase diagram as a function of the correlation strength $U/t$ and electron occupation number $N$ of the bilayer two-orbital Hubbard model for La$_3$Ni$_2$O$_7$. The physical system, indicated by the red star near $N=3$ (where $N$ counts the Ni $e_g$ electrons per unit cell) 
    resides in the bad-metal regime, situated in
    proximity to an orbital-selective Mott phase (OSMP). The black line above the solid circle denotes a Mott insulator (MI) at the (so-far not-reached) heavily-electron doped case of $N=4$. 
    Adapted from Ref.\,\cite{Liao_PRB_2026}.
    (b) Orbital-differentiated properties of the bad-metal state. The $d_{z^2}$ orbital exhibits a relatively large incoherent spectral weight, which contributes to the formation of 
    effective spin moments, whereas the $d_{x^2-y^2}$ orbital is characterized by a relatively more 
    coherent part that makes the system metallic.
    (c) 
    Illustration of the effective 
    spin moment coming from 
    the relevant $d_{z^2}$ and $d_{x^2-y^2}$ orbitals. (d) 
    The dominant 
    spin exchange interactions in the system, which emerge from an interplay between itinerant-electron-mediated RKKY coupling and 
    superexchange coupling (see the main text).} 
\label{fig:schematic_phase}
\end{figure}

In this work, we propose a theoretical framework that
is anchored by the orbital-selective correlations experimentally observed 
in the normal state~\cite{Liao_PRB_2023,Liao_PRB_2026}. The latter is illustrated in Fig.\,\ref{fig:schematic_phase}(a). 
This regime,
shown as 
bad metal,
is in proximity to an orbital selective Mott phase. 
Here, as we will describe,
 there are two distinct processes contributing to 
the exchange interactions, corresponding to 
Ruderman-Kittel-Kasuya-Yosida (RKKY) type magnetic exchange
and  superexchange interactions.
We show that an 
antiferromagnetic ground state with an ordering wavevector close to $\bm{Q}=(\pi/2,\pi/2)$ naturally develops. We demonstrate that this magnetically ordered phase supports multi-branch magnetic excitations, which are 
consistent with
the recent experimental observations. Our results underscore the essential role of orbital-selective electron correlations in 
determining the magnetic properties of the bilayer nickelate and provide direct insight into understanding the origin of high-temperature superconductivity in this system.
 
\textit{Models and Conventions.}~We describe the
La$_3$Ni$_2$O$_7$ system by a bilayer two-orbital Hubbard
model $H=H_{\rm{TB}} + H_{\rm{int}}$.
Here, the kinetic part is described by a tight-binding (TB)
Hamiltonian:
\begin{align}
    H_{\rm{TB}} &=\sum_{ijll^\prime\alpha\beta\sigma} t^{\alpha\beta}_{il,jl^\prime} d^\dagger_{il\alpha\sigma} d_{jl^\prime\beta\sigma} \, 
    \label{eq:tight-binding}
+ \sum_{il\alpha\sigma} \epsilon_\alpha d^\dagger_{il\alpha\sigma} d_{il\alpha\sigma}
\end{align}
where the 
$t^{\alpha\beta}_{il,jl^\prime}$ and $\epsilon_\alpha$ refer to hopping integrals and crystal-field levels
taken from Ref.~\onlinecite{Liao_PRB_2026}, respectively.
The local Hubbard-Kanamori interaction on each Ni site reads
\begin{align}
H_{\rm int}
&=
U\sum_{i,l,a}
n_{i l a\uparrow}n_{i l a\downarrow}
+
U'\sum_{i,l,\sigma,\sigma'}
n_{i l x\sigma}n_{i l z\sigma'}
\nonumber\\
&\quad
-J_H\sum_{i,l,\sigma,\sigma'}
d^\dagger_{i l x\sigma}
d^\dagger_{i l z\sigma'}
d_{i l x\sigma'}
d_{i l z\sigma}
\nonumber\\
&\quad
+J_H\sum_{i,l}
\left(
d^\dagger_{i l x\uparrow}d^\dagger_{i l x\downarrow}
d_{i l z\downarrow}d_{i l z\uparrow}
+{\rm H.c.}
\right),
\label{eq:kanamori}
\end{align}
where the intraorbital
Hubbard repulsion $U$, the interorbital density repulsion $U^\prime$, and the Hund's rule coupling $J_{\rm{H}}$ satisfy
the rotationally invariant relation
$U'=U-2J_H$ \cite{Castellani1978}.

Building upon the recently introduced
global phase diagram~\cite{Liao_PRB_2026}, shown in Fig.\,\ref{fig:schematic_phase}(a), the system is characterized by substantial electronic correlations. At the physical filling $N=3$, 
which counts the number of Ni $e_g$ electrons per unit cell,
it resides in a bad-metal regime with strong orbital selectivity, which 
is under the influence of an orbital-selective Mott phase (OSMP) proximate to the putative half-filled ($N=4$) Mott insulating state. In this bad-metal 
regime, the single-particle spectral function for each orbital decomposes into coherent and incoherent components --
c.f., Fig.\,\ref{fig:schematic_phase}(b).
 The coherent component lies close to the Fermi energy $E_{\rm F}$, whereas the charge excitations in the incoherent component are largely gapped, such that its dominant contribution to the low-energy
physics is an effective local moment --
as illustrated in Fig.\,\ref{fig:schematic_phase}(c).
Owing to the pronounced orbital selectivity in La$_3$Ni$_2$O$_7$, the spectral weight of the $d_{x^2-y^2}$ orbital resides 
more 
in the coherent part, while that of the $d_{z^2}$ orbital mainly populates the incoherent part. The local moment is therefore primarily associated with the $d_{z^2}$ orbital, despite both orbitals contributing to the overall spin spectral weight (see Fig.~\ref{fig:schematic_phase}(c)).

To construct a tractable low-energy description, we decompose the original hopping terms into two sectors. The first sector contains the incoherent degrees of freedom exclusively, which gives rise to superexchange interactions among the local moments. The second sector involves the coherent quasiparticles and their hybridization with the incoherent part. Integrating out the coherent fermions transforms this hybridization into an effective RKKY interaction among the local moments. To make concrete progress, and taking advantage of the contrast between the dominant weight in the incoherent and coherent parts of the spectrum from the two orbitals, we adopt a first approximation in which the coherent part is attributed solely to the $d_{x^2-y^2}$ orbital with quasiparticle weight $Z_x$, and the incoherent part to the $d_{z^2}$ orbital with weight $1-Z_z$. Values of quasiparticle spectral weights are adapted from Ref.~\cite{Liao_PRB_2026} via a slave-spin approach~\cite{Yu_PRB_2012, Yu_PRB_2017}. 
In other words, we construct the exchange interactions 
by considering the limit $Z_x \gg Z_z$, though we expect 
the results to qualitatively apply over a wider parameter regime. 
Two distinct processes contribute to the exchange coupling. One is the  superexchange coupling between the 
$d_{z^2}$-derived local moments on neighboring $\delta$-sites, which reads $J^{\rm s}_{il,jl^\prime} \sim 4(1-Z_z)^2 t_{{il,jl^\prime}}^2 / U$
\cite{Ding_PRB_2019}; this contribution, familiar in $d$-electron systems, is an adiabatic continuation of what happens in the single-orbital or degenerate-multi-orbital systems.

The other is an 
RKKY coupling,
which takes the form
$J_{
ll'}^{\text{r}}(\bm{q})=-j^2_{xz}(\bm{q})\chi_{ll'} \, , 
(\bm{q})$,
which can be considered as the adiabatic continuation of what happens 
in the extremely orbital-selective case of Anderson lattice models~\cite{WangSi_unpublished}.
Here, $j_{xz} ({\bm{q})}$ denotes a smooth $\bm{q}$-dependent inter-orbital 
exchange coupling between the $d_{z^2}$ orbital local moments and the coherent $d_{x^2-y^2}$ orbital carriers, and $\chi_{ll'}(\bm{q})$ is the layer resolved magnetic susceptibility at the wave vector $\bm{q}$ 
in 
the $d_{x^2-y^2}$ orbital channel.
We then arrive at the following effective spin Hamiltonian to describe the magnetism of the La$_3$Ni$_2$O$_7$ system:
\begin{align}
H_{\rm spin}
&=
\sum_{ij,ll^\prime}
\left(J^{\rm s}_{{il,jl^\prime}}+
J^{\rm r}_{il,jl'}
\right)
\bm S_{il}\cdot \bm S_{jl^\prime}
\label{eq:spin-model}
\end{align}
Here $\bm S_{il}$ denotes the effective local spin degree of freedom coming from
the incoherent part of the electron 
spectrum. 
Further details are provided in the Supplemental Material~\cite{SM}.

\begin{figure}
    \centering
    \includegraphics[width=.8\linewidth]{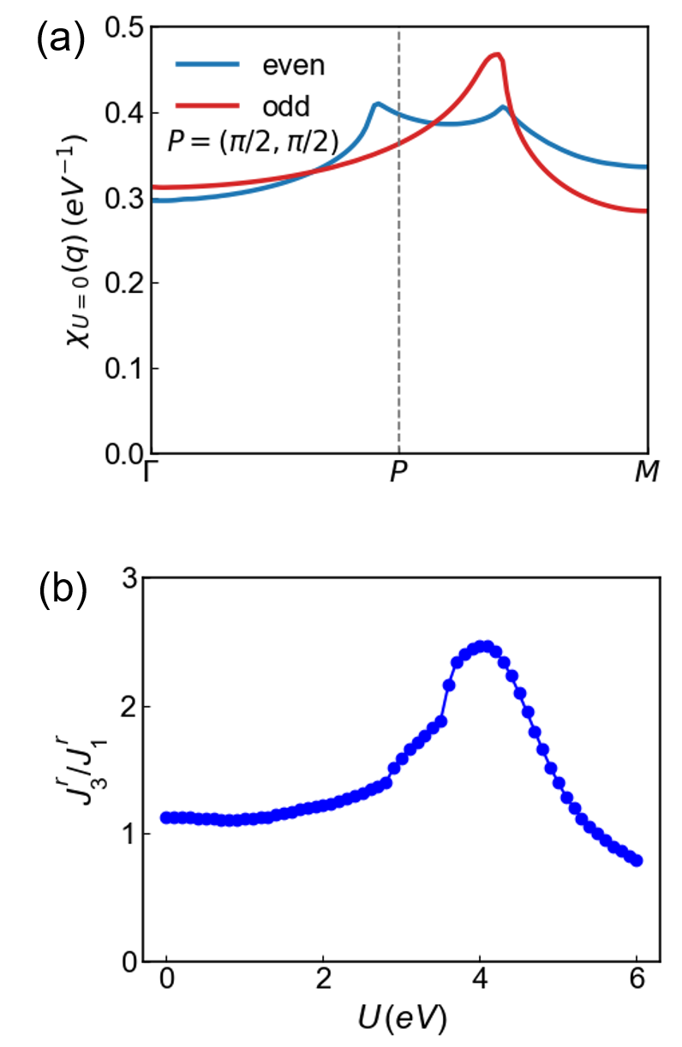}
    \caption{(a) Static spin susceptibility for $x^2-y^2$ orbitals at $N=3$ and $U=0$. (b) Ratio of the intralayer RKKY interactions $J^{\rm r}_{3}/J^{\rm r}_{1}$ as a function of $U$.
    \label{fig:chi0}}
\end{figure}

\textit{The RKKY interaction.}~Taking into account
 the mirror symmetry between the two layers $A$ and $B$, we define the susceptibilities in even and odd channels as superpositions of the layer-resolved susceptibilities:
\begin{align}
 \chi_{\rm even} &= \chi_{AA}+\chi_{BB}+\chi_{AB}+\chi_{BA}, \nonumber\\
 \chi_{\rm odd}  &= \chi_{AA}+\chi_{BB}-\chi_{AB}-\chi_{BA}.
\end{align}
Figure~\ref{fig:chi0}(a) presents the $d_{x^2-y^2}$-projected $\chi_{\rm even}(\bm{q})$ and $\chi_{\rm odd}(\bm{q})$ in the non-interacting limit ($U=0$) for wave vectors $\bm{q}$ along the $(0,0)$--$(\pi,\pi)$ direction, at the nominal electron filling $N=3$. Notably, $\chi_{\rm odd}(\bm{q})$ exhibits a prominent peak not too far away from $(0.5\pi,0.5\pi)$.
The peak arises from nesting between the electron-like $\alpha$ and hole-like $\beta$ Fermi-surface sheets. 
Although the precise value of this peak wavevector is somewhat
sensitive to the details in the tight-binding
parameterization of the DFT-derived bands, it is generically close to
exactly $(0.5\pi,0.5\pi)$; in our calculation, it is
at $\bm{q}_1\approx(0.6\pi,0.6\pi)$.
$\chi_{\rm even}(\bm{q})$ also displays several peaks at
nearby wave vectors.
As the interaction $U$ increases, the coherent quasiparticle spectral weight of the $d_{x^2-y^2}$ orbital is progressively reduced.

We now Fourier transform 
the momentum-space RKKY interaction into real space. Our analysis reveals that the dominant coupling is the third-nearest-neighbor term, $J^{\rm{r}}_3$. As shown in Fig.\,\ref{fig:chi0}(b), the ratio $J^{\rm{r}}_3/J^{\rm{r}}_1$ remains greater than unity over a broad range of $U$ and increases significantly with growing $U$, reaching its largest values near $U \sim 4$ eV. 
At the same time, the in-plane superexchange interaction will
be short-ranged. Thus, we expect the leading exchange interactions
to come from the short-range part.

\begin{figure}
    \centering
    \includegraphics[width=.8\linewidth]{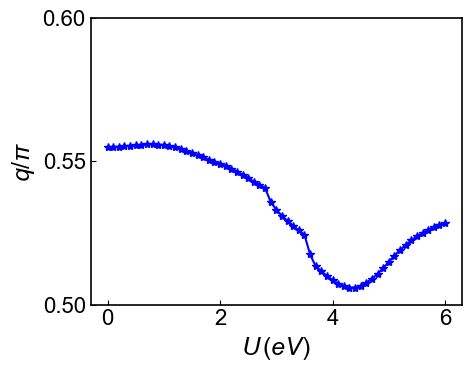}
\caption{The $U$ dependence of the component $q$ of the ordering wave vector $\bm{Q}=(q,q)$ of the magnetic ground state of an effective local spin model involving superexchange and RKKY interactions obtained via variational optimization. The wave vector 
is close to 
$\bm{Q}=(\pi/2,\pi/2)$.
\label{fig:wave vector}}
\end{figure}

\textit{Magnetic order and excitations.}~As discussed above, for 
intermediate electronic correlations ($U\sim4$ eV), the 3rd-nearest-neighbor RKKY coupling $J^{\rm{r}}_3$ 
becomes dominant, leading to a growing degree of magnetic frustration.

To explicitly determine these consequences, we analyze the magnetic properties of the spin model in Eq.~\ref{eq:spin-model}.
The model is then simplified to 
a $J^\perp$-$J_1$-$J_3$ model
on the bilayer square lattice, where $J^\perp$ is dominated by the interlayer superexchange, while $J_1$ and $J_3$ are predominantly sourced from the RKKY couplings.

For each set of exchange parameters, we first identify the candidate ordering wave vector using the Luttinger–Tisza method~\cite{LuttingerTisza_PR_1960}, which seeks the momentum-space minimum of $J(\bm{q})$. This ordering vector is subsequently refined via a variational optimization~\cite{dahlbom2025sunny}, which minimizes the classical magnetic energy with respect to both the propagation vector and the spin orientations within the reference unit cell. Within the parameter range considered, the ground state is an antiferromagnet with spins in the top and bottom layers aligned antiparallelly. In each layer, a magnetic order is stabilized at the wave vector $(q,q)$. At $U=0$, $q\approx 0.56\pi$, close to the peak position $\bm{q}_1$ of $\chi_{\rm odd}$. As $U$ increases, the enhanced ratio $J_3/J_1$ drives a gradual 
shift of the magnetic ordering vector $\bm{Q}$, as illustrated in Fig.~\ref{fig:wave vector}. This behavior 
is consistent with the 
magnetic order observed 
experimentally.

\begin{figure}
    \centering
    \includegraphics[width=.8\linewidth]{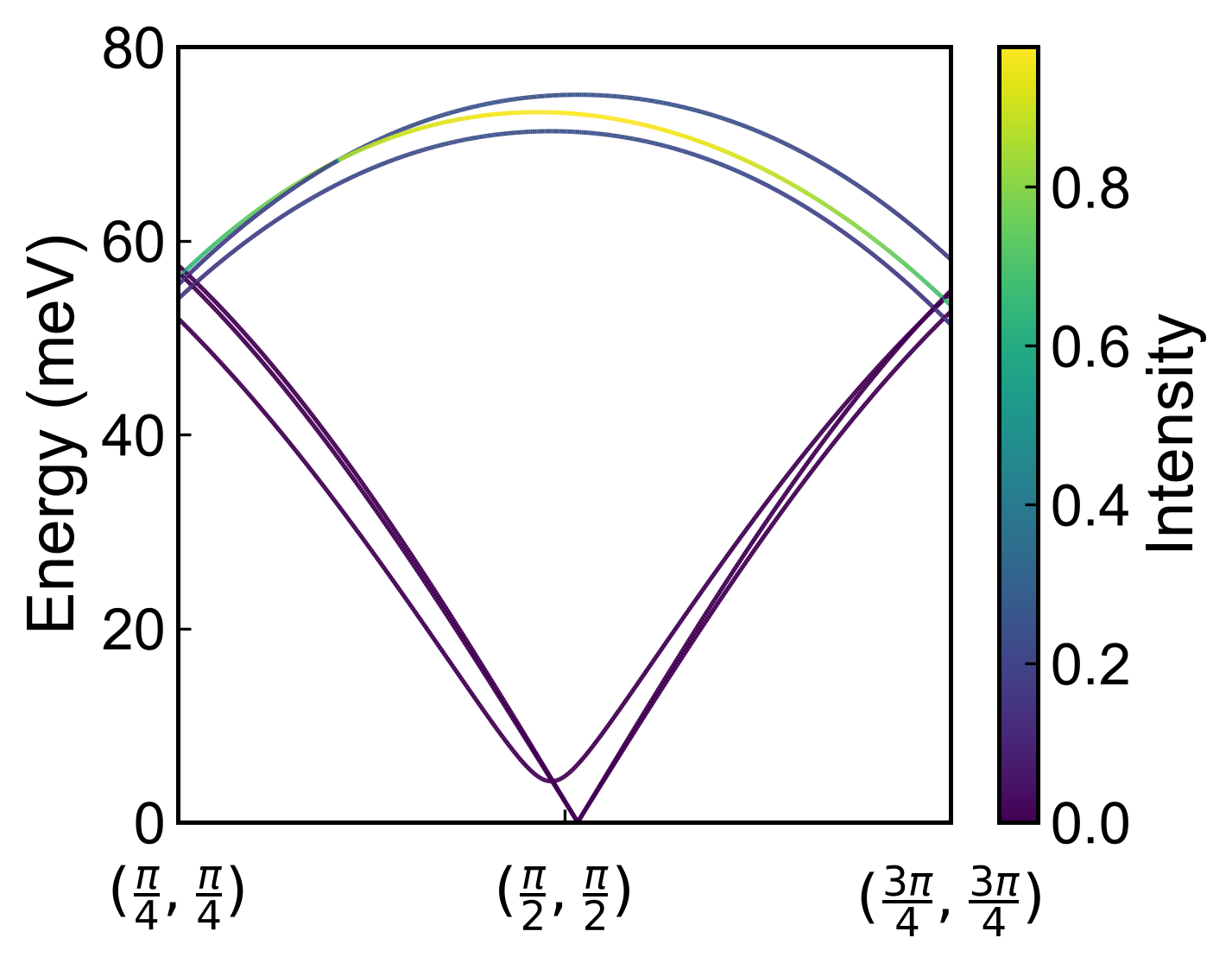}
\caption{Intensity-weighted spin-wave spectrum along $(\pi/4,\pi/2)$--$(3\pi/4,\pi/2)$ of the 2-Ni Brillouin zone in the antiferromagnetic ground state with ordering wave vector $\bm{Q}=(0.508\pi,0.508\pi)$ at $U=4.0~\mathrm{eV}$. The corresponding exchange parameters are $J^\perp S=75$ meV,
    $J_1S=1.9$ meV, $J_3S=4.6$ meV, $J_1'S=1.38$ meV (see the main text). 
 }
    \label{fig:sw}
\end{figure}

Using the optimized classical magnetic configuration as a reference, we compute the corresponding magnetic excitations. Fig.~\ref{fig:sw} presents the calculated spin-wave spectrum in the vicinity of the ordering wave vector $(\pi/2,\pi/2)$, along the $(\pi/4,\pi/4)$--$(3\pi/4,3\pi/4)$ direction of the two-Ni Brillouin zone at $U=4$ eV. With
the approach detailed in the Supplemental Material~\cite{SM}, the exchange parameters are as follows: the interlayer superexchange $J^\perp S=75$ meV, the intralayer RKKY interactions $J_1S=1.9$ meV, $J_3S=4.6$ meV, and the interlayer RKKY interaction $J_1'S=1.38$ meV. The corresponding spectrum comprises a low-energy acoustic branch and a higher-energy optical branch; the acoustic branch softens at the ordering wave vector $\bm{Q}\approx(\pi/2,\pi/2)$. Owing to the incommensurate nature of the order, each branch splits into three modes: the original mode $\epsilon(\bm{q})$ and two folded replicas $\epsilon(\bm{q}\pm\bm{Q})$. The overall bandwidth of the calculated spectrum is approximately $80$ meV, in good agreement with the energy scale observed in recent RIXS experiments~\cite{fengrixs}.

\textit{Discussions and conclusions.}~
We have shown how a magnetic order 
with the ordering wavevector near 
$(\pi/2,\pi/2)$ develops in a natural and robust way from
the strong orbital-selective correlations
observed in the normal state of the bilayer nickelates.
In addition to providing the understanding of the scattering experiments
mentioned earlier, the proposed mechanism is also consistent with a recent
systematic experiment~\cite{Wu-Shen2026}.
More broadly, the interplay between the superexchange and RKKY interactions within an orbital-selective framework serves as a conceptually new 
way to understand the magnetic correlations of multiorbital 
$d$-electron systems in such a bad metal regime.

We briefly discuss several topics that deserve further investigations. 
We have focused on the understanding of the magnetic order in bilayer nickelates;
while symmetry dictates an associated charge order,
we leave a systematic study of the
latter as a subject of 
future studies.
Experimentally, materials quality has so far limited the 
understanding of charge order in the bilayer nickelate,
in contrast to what happens in the trilayer nickelate \cite{suthar2025multiorbitalcharacterdensitywave,GimKee2025,Norman2025,zhangIntertwinedDensityWaves2020}.
Another topic of considerable interest is the interplay between
quantum spin singlet correlations among the 
adjacent interlayer 
$d_{z^2}$ orbitals within each bilayer.
Several factors need to be taken into account. 
The Hund's coupling, which is large compared to the interlayer
$d_{z^2}$ exchange coupling,
reduces the tendency towards the 
interlayer 
$d_{z^2}$ dimer singlet~\cite{Liao_PRB_2026}.
The inter-orbital exchange coupling ($j_{xz}(\bm{q})$) considered 
in the present work goes along the same direction.
Qualitatively, these combined mechanisms ultimately enable 
a magnetically ordered ground state with 
a substantial reduction of the ordered moment,
as recently observed by neutron scattering 
measurements~\cite{Chen_INS_2026}.
These qualitative considerations notwithstanding,
it will be instructive to map out the quantitative phase 
diagram that delineates the interlayer-spin-singlet and 
antiferromagnetically ordered phases; such a detailed 
study
is left for the future.

We now turn to the implications of our findings for the superconducting pairing mechanism. The calculated spin-wave spectrum reveals intense low-energy magnetic fluctuations, particularly the softening of the acoustic mode near the incommensurate ordering wave vector $\bm{Q} \approx (\pi/2,\pi/2)$. In the presence of the nonzero
coherent electron spectral weights
(with a smaller quasiparticle weight 
for $d_{z^2}$ electrons than $d_{x^2-y^2}$ electrons), 
the relatively short-range exchange interactions can drive superconducting pairing \cite{duan_arxiv2502.09195}.
Moreover, because the system resides in close proximity to the itinerant-localized crossover, tuning a non-thermal parameter --such as pressure or electron doping -- modulates 
the interplay of magnetic
correlations and quantum fluctuations and can 
serves a dual role. It weakens 
 the long-range magnetic order and allows the same electrons to
instead develop superconducting pairing.

In conclusion, we have developed a theoretical framework based on orbital-selective electronic correlations to understand the magnetic properties of the bilayer nickelate superconductor La$_3$Ni$_2$O$_7$. Motivated by the proximity to an orbital-selective Mott phase, we construct a low-energy description in which the incoherent (mostly $d_{z^2}$) degrees of freedom give rise to local moments, while the coherent (mostly $d_{x^2-y^2}$)
electrons mediate effective RKKY interactions. The resulting spin Hamiltonian, comprising both 
RKKY and superexchange couplings, naturally stabilizes 
an antiferromagnetic order with a wave vector near $(\pi/2,\pi/2)$, in 
agreement with the recent scattering 
observations in the bilayer nickelates.
The calculated spin-wave spectrum exhibits an acoustic branch softening at the ordering wave vector and an overall bandwidth of approximately 80 meV, which are consistent with RIXS and 
inelastic neutron scattering measurements.
Our results 
strongly suggest that orbital-selective correlations 
are 
an essential ingredient for the magnetism of the bilayer nickelates. Furthermore, the findings highlight the low-energy 
magnetic correlations as promising 
routes towards the superconductivity in this system, thereby expanding the general understanding of 
unconventional and high temperature superconductivity 
in strongly correlated systems.

\begin{acknowledgments}
We thank H. Chen, X. Lu, K. M. Shen, X. Wu, Y. Wu, and J. Zhan for useful discussions. Work at Rice was primarily supported by the U.S. Department of Energy, Office of Science, Basic Energy
Sciences, under Award No. DE-SC0018197. Work at Renmin University was supported in part by the National Natural Science Foundation of China (Grants No. 12334008). Q.S. acknowledge the hospitality of the Aspen Center for Physics, which is supported by NSF Grant No. PHY-2210452.
\end{acknowledgments}

\bibliographystyle{apsrev4-2}
\bibliography{nickelate}


\clearpage
\onecolumngrid 
\begin{center}
    \textbf{\large Supplemental Material for ``Magnetic Order in bilayer Ruddlesden-Popper Nickelates''}
\end{center}
\vspace{2cm}

\setcounter{equation}{0}
	\setcounter{page}{1}
	
	\setcounter{figure}{0}
	\setcounter{table}{0}
	\makeatletter
	\renewcommand{\thefigure}{S\@arabic\c@figure}
	\renewcommand{\thetable}{S\@arabic\c@table}
	
	\renewcommand{\theequation}{S\@arabic\c@equation}

\section{Orbital-selective renormalization of the coherent electronic structure}

The tight-binding Hamiltonian of the bilayer two-orbital Hubbard model introduced in Eqns.~(1) and (2) of the 
main text reads
\begin{equation}
 H_{\mathrm{TB}}
 =
 \sum_{ijll^\prime,\alpha\beta,\sigma}
 t_{il,jl^\prime}^{\alpha\beta}
 d^\dagger_{i\alpha\sigma}d_{j\beta\sigma}
 +
 \sum_{il,\alpha,\sigma}
 (\epsilon_\alpha-\mu)n_{il\alpha\sigma},
 \label{eq:sm_bare_tb}
\end{equation}
where $l,l^\prime$ are layer indices, $i,j$ label sites within each layer, and $\alpha,\beta$ denote the orbital
indices. The hopping amplitudes $t_{il,jl^\prime}^{\alpha\beta}$ refer to the hopping integral,
$\epsilon_\alpha$ denotes the crystal-field levels of the two $e_g$ orbitals, and $\mu$ is the chemical potential.

For each interaction strength $U$, we solve the two-orbital Hubbard model
within the slave-spin mean-field approach~\cite{Yu_PRB_2012, Yu_PRB_2017}. This calculation yields the
orbital-dependent quasiparticle spectral weights $Z_\alpha (U)$ and static energy
shifts $\lambda_\alpha (U)$. We then use these quantities to construct a
renormalized tight-binding Hamiltonian for the coherent part of electrons, in which the hopping amplitudes and
onsite energies are modified according to the slave-spin results.

The resulting quasiparticle Hamiltonian is
\begin{align}
 H_{\mathrm{eff}}
 ={}
 \sum_{ijll^\prime,\alpha\beta,\sigma}
 \sqrt{Z_\alpha(U)Z_\beta(U)}\,
 t_{il,jl^\prime}^{\alpha\beta}
 d^\dagger_{il\alpha\sigma}d_{jl^\prime\beta\sigma} +
 \sum_{il,\alpha,\sigma}
 \left[
 \epsilon_\alpha+\lambda_\alpha(U)-\mu(U)
 \right]
 n_{il\alpha\sigma}.
 \label{eq:sm_qp_hamiltonian}
\end{align}
Thus, the slave-spin quasiparticle weights directly renormalize the bare
hopping amplitudes to
\begin{equation}
 \widetilde t_{il,jl^\prime}^{\alpha\beta}(U)
 =
 \sqrt{Z_\alpha(U)Z_\beta(U)}\,
 t_{il,jl^\prime}^{\alpha\beta}.
 \label{eq:sm_z_renormalization}
\end{equation}
The orbital-dependent shifts $\lambda_\alpha(U)$ similarly renormalize the
onsite energies. In this way, the correlation effects to the coherent electrons
are incorporated into an effective
quasiparticle tight-binding Hamiltonian.
This procedure incorporates the orbital-dependent quasiparticle
renormalization induced by local electronic correlations into the effective
single-particle Hamiltonian. It therefore allows us to describe the
strongly correlated and orbital-selective regime beyond a weak-coupling
treatment based on the bare tight-binding bands.

The resulting orbital-selective regime admits a simple low-energy
interpretation. The more dispersive $d_{x^2-y^2}$ quasiparticles form the
itinerant sector and determine the particle-hole susceptibility. The more
strongly renormalized $d_{z^2}$ sector mainly contributes local moment degree of freedom in low energy, whose dominant
short-range interaction is the antiferromagnetic interlayer superexchange
$J^{\perp}$. A smooth $\bm{q}$-dpendent interorbital exchange coupling $j_{xz}(\bm{q})$, taking into account effects of both onsite Hund's rule interaction and inter-site hybridizations, connects the coherent and incoherent sectors, allowing the itinerant
$d_{x^2-y^2}$ quasiparticles to mediate RKKY interactions between the
$d_{z^2}$ moments.

To evaluate the itinerant-electron response, we use the renormalized
quasiparticle Hamiltonian obtained from the slave-spin calculation in the
preceding section. On an $N_k\times N_k$ momentum mesh, we diagonalize its
single-particle matrix as
\begin{equation}
    \sum_b \left(H_{\mathrm{eff}}\right)_{ab}(\bm k;U)
    u_{bn}(\bm k;U)
    =
    \xi_n(\bm k;U)
    u_{an}(\bm k;U),
    \label{eq:sm_qp_eigenproblem}
\end{equation}
Here orbital and layer indices are combined into the indices $a,b$ (\textit{e.g.}, $a=l\alpha$).

The static susceptibility of the renormalized quasiparticles is then calculated via
\begin{align}
    \chi_{ab;cd}(\bm q;U)
    ={}&
    -\frac{1}{N_k^2}
    \sum_{\bm k}\sum_{m,n}
    u_{cm}(\bm k+\bm q;U)
    u^*_{am}(\bm k+\bm q;U)
    \nonumber\\
    &\times
    u_{bn}(\bm k;U)
    u^*_{dn}(\bm k;U)
    \mathcal{L}_{mn}(\bm k,\bm q;U),
    \label{eq:sm_chi0}
\end{align}
where
\begin{equation}
    \mathcal{L}_{mn}(\bm k,\bm q;U)
    =
    \frac{
        f[\xi_m(\bm k+\bm q;U)]
        -
        f[\xi_n(\bm k;U)]
    }{
        \xi_m(\bm k+\bm q;U)
        -
        \xi_n(\bm k;U)
    }.
    \label{eq:sm_lindhard_kernel}
\end{equation}
For vanishing denominator in the calculation, the ratio is evaluated using its limiting
value given by the derivative of the Fermi function. The interaction strength $U$ affects the susceptibility through the
slave-spin renormalized quasiparticle energies and eigenvectors.

With strong orbital selectivity, the local moments originate predominantly from the $d_{z^2}$ sector,
whereas long-range RKKY interactions are mediated by the more itinerant
$d_{x^2-y^2}$ quasiparticles. We therefore project the two susceptibility
vertices onto the $d_{x^2-y^2}$ orbital, 
and the layer-resolved susceptibility after the projection is defined as
\begin{equation}
    \chi_{ll'}(\bm q;U)
    \equiv
    \chi_{l x,l x;l' x,l' x}(\bm q;U),
    \label{eq:sm_layer_chi}
\end{equation}
where $x$ denotes the $d_{x^2-y^2}$ orbital and $l,l'=A,B$.

The even and odd bilayer channels used in the main text are then given by
\begin{align}
    \chi_{\mathrm{even}}(\bm q;U)
    ={}&
    \chi_{AA}(\bm q;U)
    +
    \chi_{BB}(\bm q;U)
    +
    \chi_{AB}(\bm q;U)
    +
    \chi_{BA}(\bm q;U),
    \nonumber\\
    \chi_{\mathrm{odd}}(\bm q;U)
    ={}&
    \chi_{AA}(\bm q;U)
    +
    \chi_{BB}(\bm q;U)
    -
    \chi_{AB}(\bm q;U)
    -
    \chi_{BA}(\bm q;U).
    \label{eq:sm_even_odd}
\end{align}

\section{RKKY exchange interactions}

In the orbital-selective regime, charge fluctuations in the
$d_{z^2}$ sector are strongly suppressed, and their low-energy
degrees of freedom are represented by localized moments.
According to the discussion in the main text, the exchange interactions among these local moments are from two distinct processes.
One is the superexchange interaction between the local moments. Based on analysis in Ref.~\onlinecite{Ding_PRB_2019}, 
the superexchange coupling for the incoherent local moments reads $J^{\rm s}_{il,jl^\prime} \sim 4(1-Z_z)^2 t_{{il,jl^\prime}}^2 / U$, which is renormalized by the spectral weights of the incoherent electrons.  

The other process contributing to exchange interaction is the RKKY interaction among the local moments mediated by the coherent electrons. The coupling between the incoherent local moments in the $d_{z^2}$ orbital and the coherent quasiparticles in the $d_{x^2-y^2}$ orbital includes both an onsite Hund's rule coupling and an inter-site hybridization. These give rise to the following interorbital exchange interaction:
\begin{equation}
    H_{xz}
    =
    \sum_{ij,l l^\prime,\bm{q}}
    j^{xz}_{ll^\prime}(\bm{q})
    \bm{S}_{il}
    \cdot
    \bm s_{jl^\prime} e^{i\bm{q}\cdot (\bm{R}_i-\bm{R}_j)},
    \label{eq:sm_hund}
\end{equation}
where $\bm S^{z}_{il}$ denotes the localized moment associated with
the $d_{z^2}$ orbital, $\bm s_{il}$ is the spin density of the
itinerant $d_{x^2-y^2}$ quasiparticles, and $j^{xz}_{ll^\prime}(\bm{q})$ is a smooth $\bm{q}$-dependent interorbital exchange coupling between $d_{x^2-y^2}$ and $d_{z^2}$ orbitals.

To the second order in $j^{xz}_{ll^\prime}(\bm{q})$, integrating out the itinerant quasiparticles
generates the effective RKKY interaction among local moments
\begin{equation}
    H_{\rm{r}}
    =
    \sum_{ij,ll'}
    J^{\rm{r}}_{ll'}(\bm R_i-\bm R_j)
    \bm S_{il}
    \cdot
    \bm S_{jl'},
    \label{eq:sm_rkky_hamiltonian}
\end{equation}
with
\begin{equation}
    J^{\rm{r}}_{ll'}(\bm R)
    =
    -\sum_{\bm{q}} e^{i\bm{q}\cdot \bm{R}} \left(j^{xz}_{ll^\prime}(\bm{q})\right)^2
    \operatorname{Re}
    \chi_{ll'}(\bm q),
    \label{eq:sm_rkky}
\end{equation}
where the momentum sum is performed over an $N_q\times N_q$ mesh.  

\begin{figure}
    \centering
    \includegraphics[width=0.5\linewidth]{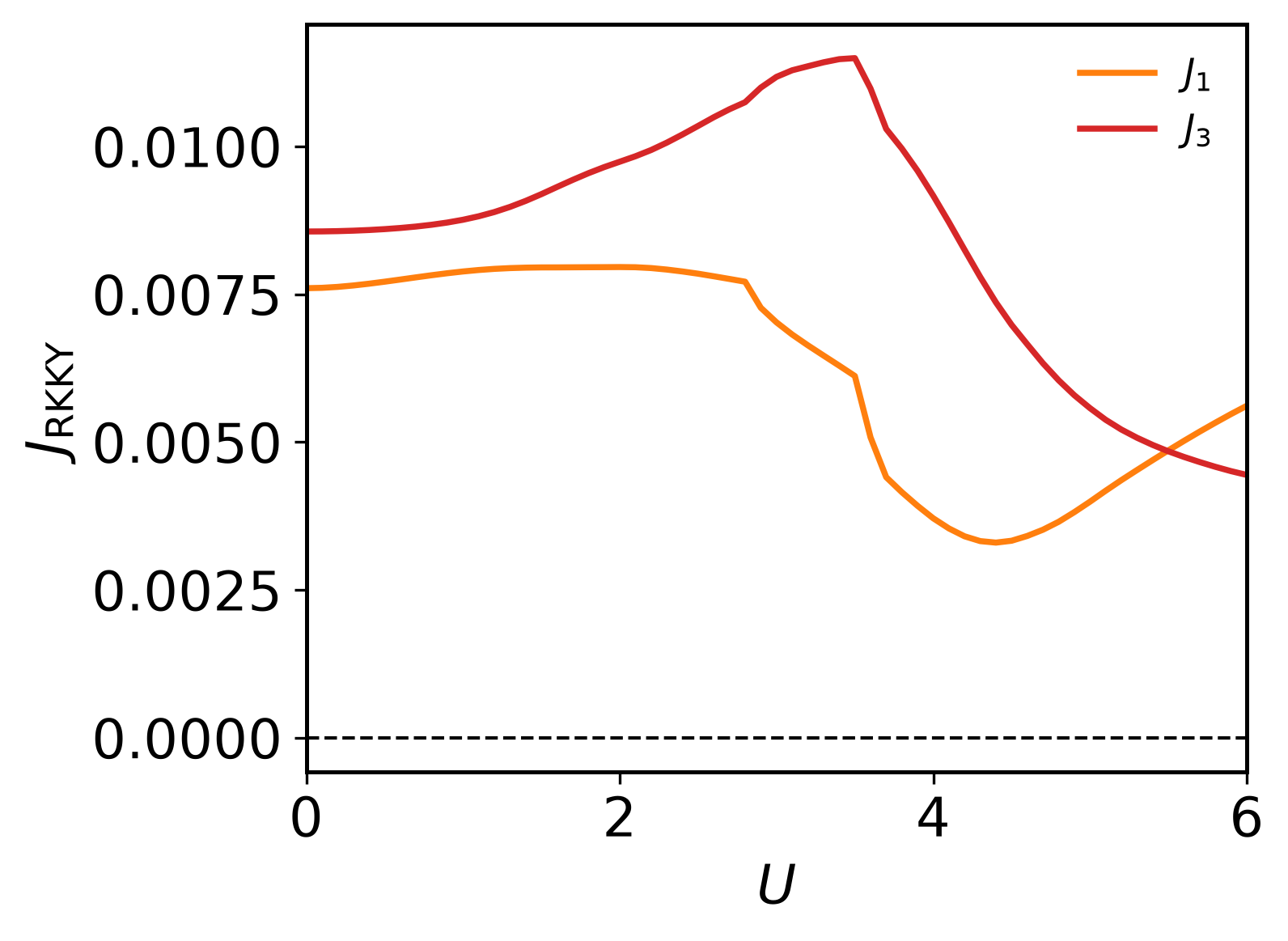}
    \caption{Calculated nearest-neighbor ($J_1$) and third-neighbor ($J_3$) RKKY exchange couplings as functions of the Hubbard interaction $U$. Both exchange couplings and $U$ are in units of eV.} 
    \label{sfig:J_vs_U}
\end{figure}

With consideration of both superexchange and RKKY interactions, we obtain the effective local spin model
for the magnetism of La$_3$Ni$_2$O$_7$ as 
\begin{equation}
    H_{\mathrm{spin}}
    =
    \sum_{ijl l^\prime}
    J_{il,jl^\prime}\,
    \bm S_{il}\cdot\bm S_{jl^\prime} = \sum_{ijl l^\prime}
    (J^{\rm s}_{il,jl^\prime} + J^{\rm r}_{il,jl^\prime})\,
    \bm S_{il}\cdot\bm S_{jl^\prime},
\end{equation}
which arrives at Eq.~(3) of the main text. For $d_{z^2}$ orbital, the interlayer nearest neighbor hopping $t^\perp$ dominant. We then expect that the corresponding interlayer exchange coupling $J^\perp=J_{iA,iB}$ is dominant by the superexchange process, \textit{e.g.} $J^\perp\approx J^{\rm s}_{iA,iB}$. With $t^\perp\sim0.6$ eV, and the $Z_z$ value from slave-spin calculation~\cite{Liao_PRB_2026}, we obtain $J^{\rm s}_{iA,iB}S \approx 75$ meV. 
On the other hand, intralayer and further neighboring interlayer couplings are estimated to be at or less than the order of 1 meV. Therefore, these exchange couplings are predominantly from RKKY interaction, \textit{e.g.} $J_{il,jl^\prime}\approx J^{\rm r}_{il,jl^\prime}$ for $i\neq j$. 
Defining the intra- and inter-layer susceptibilities as
\begin{align}
    \chi_{\rm intra}(\bm q)
    &=
    \frac{
        \chi_{AA}(\bm q)
        +
        \chi_{BB}(\bm q)
    }{2} = \frac{
        \chi_{\rm even}(\bm q)
        +
        \chi_{\rm odd}(\bm q)
    }{4},
    \nonumber\\
    \chi_{\rm inter}(\bm q)
    &=
    \frac{
        \chi_{AB}(\bm q)
        +
        \chi_{BA}(\bm q)
    }{2} = \frac{
        \chi_{\rm even}(\bm q)
        -
        \chi_{\rm odd}(\bm q)
    }{4},
    \label{eq:sm_intra_inter_chi}
\end{align}
The exchange couplings of several neighboring spin pairs are as follows:
\begin{align}
    J_1
    &=
    J^{\rm{r}}_{\rm{intra}}\left(\bm{R}_{ij}=(1,0)\right),
    &
    J_2
    &=
    J^{\rm{r}}_{\rm{intra}}\left(\bm{R}_{ij}=(1,1)\right),
    \nonumber\\
    J_3
    &=
    J^{\rm{r}}_{\rm{intra}}\left(\bm{R}_{ij}=(2,0)\right),
    &
    J^\prime_1
    &=
    J^{\rm{r}}_{\rm{inter}}\left(\bm{R}_{ij}=(1,0)\right).
    \label{eq:sm_exchange_definitions}
\end{align}
Here $\bm{R}_{ij}$ are defined on the square lattice in units of the lattice spacing. Values of these exchange couplings are obtained via Eq.~\eqref{eq:sm_rkky} by taking $|j^{xz}_{ll^\prime}(\bm{q})|\approx j^{xz}= 1$ eV (neglecting its slow $\bm{q}$ dependence, given that it is dominant by the Hund's rule coupling). Fig.~\ref{sfig:J_vs_U} shows evolution of the in-plane nearest and the 3rd-nearest neighbor exchange interactions with $U$. We find $J_3>J_1$ with increasing $U$ up to $U\lesssim 6$ eV, and the $J_3/J_1$ ratio develops a sharp peak at about $4$ eV. 

\end{document}